\documentclass[preprint,12pt,3p]{elsarticle}

\usepackage{amsmath,amssymb,amsfonts}
\usepackage{mathrsfs}
\usepackage{chemformula}
\usepackage{color}
\usepackage[normalem]{ulem}
\usepackage{stmaryrd}
\usepackage{diagbox}
\usepackage[parfill]{parskip}
\usepackage{hyperref}              %creates hyperlinks
\usepackage{float}                 % allows H option on floats to force here placement
\graphicspath{{./figures/}}
 \biboptions{sort&compress}

\newcommand{\blue}[1]{\textcolor{black}{#1}}

\newcommand{\overbar}[1]{\mkern 1.5mu\overline{\mkern-1.5mu#1\mkern-1.5mu}\mkern 1.5mu}
\newcommand{\beq}{\begin{equation}}
\newcommand{\eeq}{\end{equation}}
\newcommand{\pfr}[2]{\ensuremath{\frac{\partial #1}{\partial #2}}}
\newcommand{\pfi}[2]{\ensuremath{{\partial #1}/{\partial #2}}}
\newcommand{\sd}[2]{\frac{\partial^2 #1}{\partial #2^2}}
\newcommand{\ep}{\epsilon}

\newcommand\Pec{\mbox{\textit{Pe}}}

\newcommand\Lew{\mbox{\textit{Le}}}

\newcommand\Dam{\mbox{\textit{Da}}}

\journal{Combustion and Flame}

\begin{document}

\begin{frontmatter}

\title{Stability of diffusion flames under shear flow: Taylor dispersion and the formation of flame streets}

\author{Prabakaran Rajamanickam, Aiden Kelly, Joel Daou}
\address{Department of Mathematics, University of Manchester, Manchester M13 9PL, UK}

\begin{abstract}
Diffusion flame streets, observed in non-premixed micro-combustion devices, align parallel to a shear flow. They are observed to occur in mixtures with high Lewis number ($\Lew$) fuels, provided that the flow Reynolds number, or the Peclet number $\Pec$,  exceeds a critical value. The underlying mechanisms behind these observations have not yet been fully understood. In the present paper, we identify the coupling between diffusive-thermal instabilities and Taylor dispersion as a mechanism which is able to explain the experimental observations above. The explanation is largely based on the fact that Taylor dispersion enhances all diffusion processes in the flow direction, leading effectively to anisotropic diffusion with an effective (flow-dependent) Lewis number in the flow direction which is  proportional to  $1/\Lew$ for  $\Pec \gg 1$. Validation of the identified mechanism is demonstrated within a simple model by investigating the stability of a planar diffusion flame established parallel to a plane Poiseuille flow in a narrow channel. A linear stability analysis, leading to an eigenvalue problem solved numerically, shows that cellular (or finite wavelength) instabilities emerge for high Lewis number fuels when the Peclet number exceeds a critical value. Furthermore, for Peclet numbers below this critical value, longwave instabilities with or without time oscillations are obtained. Stability regime diagrams are presented for illustrative cases in a $\Lew$-$\Pec$ plane where various instability domains are identified. Finally, the linear analysis is supported and complemented by time dependent numerical simulations, describing the  evolution of unstable diffusion flames. The simulations demonstrate the existence of stable cellular structures and  show that the longwave instabilities are conducive to flame extinction.
\end{abstract}


\end{frontmatter}

\section{Introduction}
Formation of diffusion flame streets in non-premixed micro-combustion channels, first reported in~\cite{miesse2004submillimeter,miesse2005experimental,miesse2005diffusion}, has been shown to exhibit peculiar features that are not observed in large scale burners. First of all, they are found to occur readily when using heavier fuels  but not lighter fuels such as hydrogen. In the latter case, a continuous diffusion flame is observed~\cite{miesse2004submillimeter,miesse2005diffusion} unless the inlet mixture is sufficiently diluted with helium which is known to increase the   effective Lewis number. In summary, flame streets are found to occur in non-premixed combustion when the effective  Lewis number is sufficiently large~\cite{miesse2005diffusion}. Moreover, they are observed only if the flow Reynolds number (or, the flow rate) exceeds a critical value~\cite{miesse2004submillimeter,miesse2005experimental,miesse2005diffusion,prakash2007flame,xu2009studies}; sufficiently above this critical value the streets are found to be stable. When the Reynolds number is lower than the critical Reynolds number, typically the quenched state is observed. However, depending on the conditions prevailing at the channel exit, sometimes repetitive extinction-ignition events starting from the exit region and involving edge flame propagation are observed. Computational studies investigating these flame streets~\cite{mohan2017diffusion,mackay2019steady,kang2019numerical,sun2020numerical} have qualitatively reproduced the experimental observations. 

The critical Reynolds number can depend on a number of relevant controlling parameters such as the mixture strength, heat loss mechanism, type of the channel cross-section, etc; for example, the critical Reynolds number is found to decrease with increasing wall temperature~\cite{xu2009studies}. Regardless, the important point to note is that the existence of a critical Reynolds number provides strong evidence for a flow-induced effect causing the formation of a flame street. However, the underlying physical mechanisms behind the experimental observations have not been yet fully identified. In particular, despite numerous efforts, the following two issues still lack clear explanations:
\begin{enumerate}
    \item Why planar diffusion flames aligned with the shear flow do not exist for heavier fuels when effective Lewis number of the inlet mixtures is large?
    \item What justifies the existence of a critical Reynolds number (or, equivalently a critical Peclet number) above which stable diffusion flame streets are encountered?
\end{enumerate}
\vspace{-0.4cm}
The suggestion in \cite{miesse2005experimental,miesse2005diffusion} that the observed patterns are the result of an instability appears not to have been pursued yet. In the present contribution, we shall show that the two aforementioned issues may be explained as being caused by well-known diffusive-thermal flame instabilities, provided account is made of Taylor's dispersion mechanism~\cite{taylor1953dispersion}.

Although Taylor dispersion played a fundamental role in a wide range of practical applications involving unidirectional flows, its importance in combustion systems has not been recognized until recently. Its influence on the structure and stability of premixed flames are investigated in~\cite{pearce2014taylor,daou2018taylor,rajamanickam2023thick,daou2021effect,daou2023flame} and the structure of Burke-Schumann diffusion flame is studied in~\cite{linan2020taylor,rajamanickam2022effects} . An important result revealed by the analyses of~\cite{daou2018taylor,linan2020taylor} is the presence of a\emph{ flow-dependent effective Lewis number} in the flow direction, say the  $x$-direction, given by
\begin{equation}
     \Lew_{x}=\frac{\Lew(1+\gamma \Pec^2)}{1+\gamma \Pec^2\Lew^2}.  \label{Lex}
\end{equation}
Here $\Lew$ is the fuel Lewis number, $\Pec$ the Peclet number and $\gamma$   a constant which is determined by the velocity profile. The significance of this formula is clearly elucidated by evaluating it for the following three values of $\Pec$:
\begin{align}
    \Pec=0\,\Rightarrow\, \Lew_x = \Lew, \qquad  \Pec=1/\sqrt{\gamma\Lew}\,\Rightarrow\, \Lew_x = 1, \qquad  \Pec\gg 1\,\Rightarrow\,  \Lew_x = 1/\Lew. \label{Lex asym}
\end{align}
 Thus, the flow-dependent effective Lewis number, $\Lew_x$, varies from being purely determined by molecular diffusion and equal to $\Lew$ in the absence of flow to being equal to $1/\Lew$ when $\Pec\gg 1$.  In other words, the effect of a shear flow is such that a weakly diffusing scalar appears effectively as a strongly diffusing scalar and a strongly diffusing scalar appears effectively as a weakly diffusing one if $\Pec>1/\sqrt{\gamma\Lew}$. This peculiar feature of Taylor dispersion is expected to produce interesting effects on flame characteristics. For instance, when the diffusion flame lies perpendicular to the fluid stream~\cite{linan2020taylor}, subadiabatic flame temperatures are predicted for low Lewis number fuels and superadiabatic flame temperatures for large Lewis number fuels if $\Pec>1/\sqrt{\gamma\Lew}$. Although formula~\eqref{Lex} is not strictly valid when thermal expansion effects are taken into account, the two limiting forms in~\eqref{Lex asym} for $\Pec=0$ and $\Pec\gg 1$ are still correct as shown in~\cite{rajamanickam2022effects}. Finally, we can note from formula~\eqref{Lex} that $\Lew_x=1$ for all values of $\Pec$ if $\Lew=1$. %this is so because for unity Lewis number cases, Taylor dispersion alters the mass diffusivity and thermal diffusivity by an equal amount such that their ratio remains unity.

 \begin{figure}[h!]
\vspace{-2cm}
\centering
\includegraphics[scale=0.7]{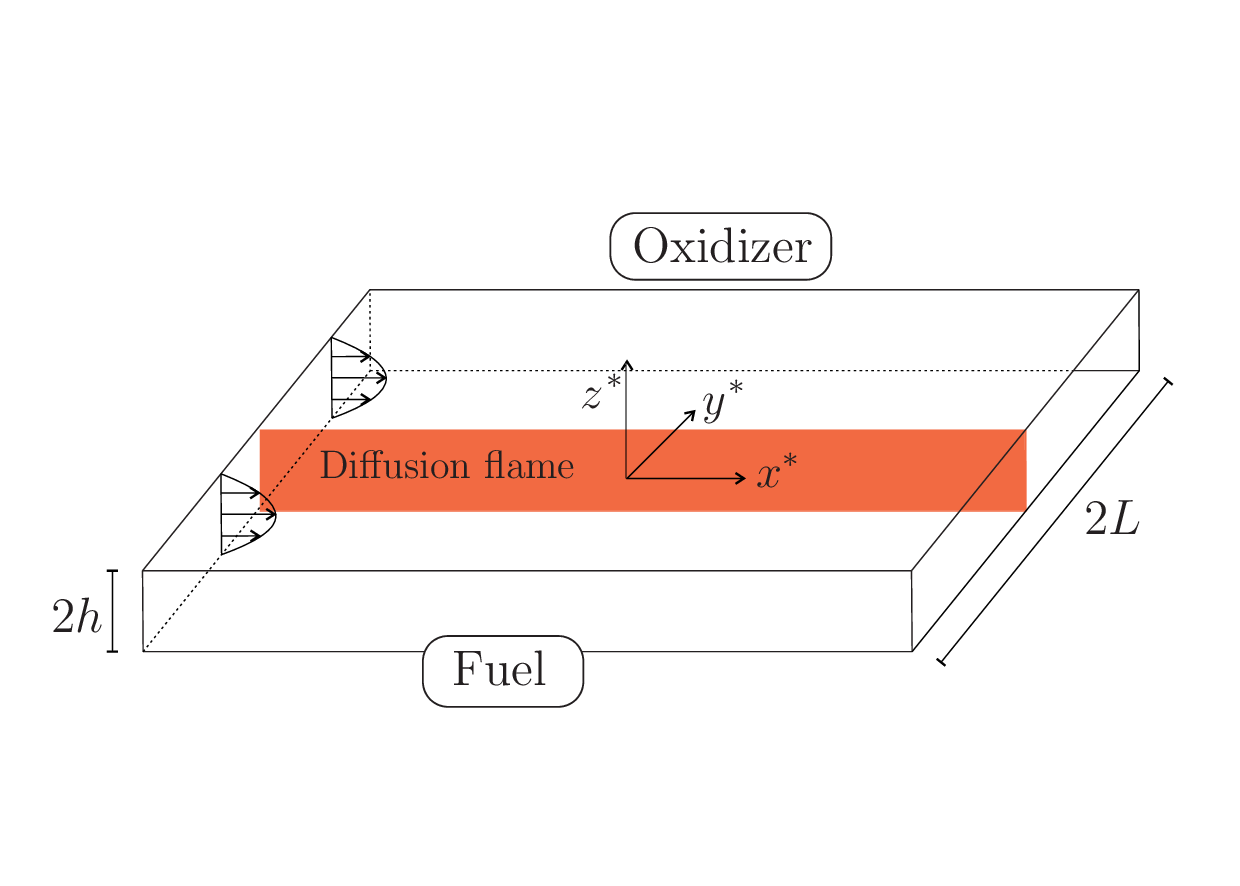}
\vspace{-1.5cm}
\caption{Model configuration showing a diffusion flame aligned parallel to a shear (Poiseuille) flow. The fuel and oxidizer reach the flame by transverse diffusion.}
\label{fig:Setup}
\end{figure}

It is important to emphasize that the enhancement of diffusion is in the longitudinal (or, flow direction) only and not in the transverse direction. This leads to anisotropic diffusion characterised by $\Lew_x$ in the $x$-direction and by $\Lew$ in the transverse direction. It follows that the flame alignment with respect to the flow is an essential aspect to consider when attempting to explain the observations in nonpremixed micro-combustion devices. %Further remarks about flame alignment and flame instabilities are provided in $\S$6.

In the present paper, we shall show that the issues mentioned above can be explained as being the consequence of diffusional-thermal instabilities occurring in a medium where diffusion is effectively anisotropic due to Taylor dispersion. This is demonstrated by investigating the stability of a planar diffusion flame aligned with the shear flow accounting for the presence of effectively two distinct Lewis numbers, $\Lew_x$ and $\Lew$, in the longitudinal and transverse directions respectively. \blue{It is important to emphasize that this problem, which addresses the stability  of the diffusion flame under the influence of shear-enhanced diffusion, is both novel and fundamental and therefore deserves investigation on its own right. To this end, and in order to highlight the new instability mechanism involved, we shall purposely adopt suitable simplifying assumptions. These include ignoring complications related to entrance and exit conditions by focusing on the fully-developed flow region as well as ignoring other effects such as the presence of heat losses and density variations.} The model configuration is sketched in Fig.~\ref{fig:Setup} which depicts a diffusion flame aligned parallel to a two-dimensional Poiseuille flow. The instabilities of the underlying one-dimensional diffusion flames have been studied previously,  in the absence of longitudinal flows, in~\citep{kirkbey1966analytical,kim1996diffusional,kim1997linear,sohn1999instability,sohn2000nonlinear,mackay2019steady}.

The paper is structured as follows. The problem is formulated in $\S$2. The linear stability of the planar diffusion flame is characterized by an eigenvalue problem derived in $\S$3. The solution of the eigenvalue problem is presented in $\S$4, where the emerging instabilities are discussed and delineated in a $\Lew$-$\Pec$ plane. Numerical simulations complementing the linear stability findings are presented in $\S$5 for selected cases, illustrating the nonlinear time evolution of unstable flames.

\section{Governing equations and boundary conditions}
We consider a simple constant-density model to investigate the stability of a planar flame aligned with the direction of a shear flow, as sketched in Fig.~\ref{fig:Setup}. Thermal expansion and heat loss effects are ignored so as to demonstrate that account of Taylor dispersion is sufficient to obtain the instabilities under consideration.   The chemistry is modeled by a single step irreversible Arrhenius reaction for which the mass of fuel burnt per unit volume per unit time is given by
\begin{equation}
    B \rho^2 Y_F^* Y_O^* \exp(-E/RT^*).   \nonumber
\end{equation}
Here $B$ is the preexponential factor, $\rho$ is the constant density, $Y_i^*$ is the mass fraction of species $i$, $T^*$ is the temperature, $E$ is the activation energy and $R$ is the universal gas constant. 

The temperatures at both sides $y=\pm L$ are assumed  equal and denoted by $T_u$. \blue{It is also assumed that the concentrations of fuel and oxidizer are maintained fixed at $y=\pm L$, namely, $Y_F^*=Y_{F,F}$, $Y_O^*=0$ at the fuel side $y=-L$ and $Y_F^*=0$, $Y_O^*=Y_{O,O}$ at the oxidizer side $y=+L$. Although such conditions may be delicate to fully achieve experimentally~\cite{robert2012thermal,robert2013experimental}, they are adopted herein for sake of simplicity, as it is done in many other theoretical investigations involving diffusion flames such as~\cite{kim1996diffusional,kim1997linear,sohn1999instability,sohn2000nonlinear,mackay2019steady,buckmaster2000holes,daou2010triple,pearce2013rayleigh,law2010combustion,daou2019flame,vance2001stability}.} 

The Zeldovich number $\beta$, the heat release parameter $\alpha$, the stoichiometry parameter $S$ and the Damk\"{o}hler number $\Dam$ are defined by
\begin{equation}
    \beta = \frac{E(T_{ad}-T_u)}{RT_{ad}^2} ,\,\, \alpha = \frac{T_{ad}-T_u}{T_{ad}}, \,\, S = \frac{sY_{F,F}}{Y_{O,O}}, \,\, \Dam = \frac{L^2/D_T}{\beta^3[B\rho Y_{O,O}\exp(-E/RT_{ad})]^{-1}} \label{Dam}
\end{equation}
where $T_{ad}= T_u + qY_{F,F}/c_p(S+1)$ is the adiabatic flame temperature at the stoichiometric location. Further, $s$ and $q$ denote the mass of oxygen consumed and the amount of heat released per unit mass of fuel burnt, $c_p$ the constant-pressure specific heat and $D_T$ the constant thermal diffusivity. The oxidizer Lewis number $\Lew_O$ is taken equal to unity and the fuel Lewis number $\Lew_F$ is  simply denoted as $\Lew$.

The problem is analyzed in a frame of reference moving with the flow mean velocity $U$. In this frame, the velocity components are given by
\begin{equation}
    v_x = \frac{U}{2}\left(1-\frac{3z^{*2}}{h^2}\right), \quad v_y=v_z=0.  \nonumber
\end{equation}
In terms of the non-dimensional variables
\begin{align}
   t= \frac{t^*}{L^2/D_T}, \quad (x,y,z) = \frac{1}{L}(x^*, y^*,z^*/\ep), \quad Y_F = \frac{Y_F^*}{Y_{F,F}}, \quad Y_O = \frac{Y_O^*}{Y_{O,O}}, \quad \theta = \frac{T^*-T_u}{T_{ad}-T_u} \nonumber
\end{align}
the governing equations read
\begin{align}
  \pfr{Y_F}{t} +  \frac{\Pec}{2\ep} (1-3z^2) \pfr{Y_F}{x} &= \frac{1}{\Lew} \left(\pfr{^2Y_F}{x^2} + \pfr{^2Y_F}{y^2} + \frac{1}{\ep^2} \pfr{^2Y_F}{z^2}\right) -  \Dam\, \omega  \label{yF}\\
 \pfr{Y_O}{t} +  \frac{\Pec}{2\ep} (1-3z^2)  \pfr{Y_O}{x} &=  \pfr{^2Y_O}{x^2} + \pfr{^2Y_O}{y^2} +  \frac{1}{\ep^2}\pfr{^2Y_O}{z^2} - S \Dam \, \omega \label{yO}\\
   \pfr{\theta}{t} + \frac{\Pec}{2\ep} (1-3z^2) \pfr{\theta}{x} &= \pfr{^2\theta}{x^2} + \pfr{^2\theta}{y^2} +  \frac{1}{\ep^2}\pfr{^2\theta}{z^2} + (S+1) \Dam\, \omega, \label{theta}
\end{align}
where $\ep=h/L$, $\Pec=Uh/D_T$ and
\begin{equation}
   \omega = \beta^3  Y_FY_O \exp\left[\frac{\beta(\theta-1)}{1+\alpha(\theta-1)}\right]. \label{omega}
\end{equation}

In the limit $\ep\rightarrow 0$, the dependent variables tend to be uniform in the $z$ direction and the convective terms on the left-side of~\eqref{yF}-\eqref{theta} lead upon depth averaging of the equations to enhanced longitudinal diffusion as derived in~\cite{pearce2014taylor,daou2018taylor,linan2020taylor,rajamanickam2022effects}. The resulting problem is two dimensional and is given by
\begin{align}
  \pfr{Y_F}{t}  &= \frac{1}{\Lew} \left[(1+p^2\Lew^2)\sd{Y_F}{x} + \sd{Y_F}{y} \right] - \Dam\, \omega \label{yF nondim}\\
  \pfr{Y_O}{t}  &= (1+p^2)\sd{Y_O}{x} + \sd{Y_O}{y}  - S  \Dam \, \omega\label{yO nondim}\\
   \pfr{\theta}{t}  &= (1+p^2)\sd{\theta}{x} + \sd{\theta}{y}  + (S+1)\Dam \, \omega,\label{theta nondim}
\end{align}
where $p = \sqrt{\gamma} \Pec$ and $\gamma = 2/105$. The corresponding lateral boundary conditions are
\begin{align}
    &Y_F = 1, \quad Y_O = 0, \quad \theta = 0 \quad \text{at} \quad y=-1,\label{nondim BC1}\\
     &Y_F = 0, \quad Y_O = 1, \quad  \theta = 0 \quad \text{at} \quad y=+1. \label{nondim BC2}
\end{align}
\blue{As for the boundary condition in the $x$-direction, we shall adopt, for simplicity, periodic conditions, which will allow us in particular to avoid complications associated with entrance and exit conditions. This is sufficient for our main purpose, which is to identify the coupling between shear-enhanced diffusion and the instability of the planar diffusion flame, the focus being on the region where the flow is fully developed. For investigations that incorporate the entrance region conditions and other effects such as heat loss, the reader is referred to~\cite{mohan2017diffusion}, where a two-dimensional depth-averaged model is derived and solved numerically, and to~\cite{mackay2019steady,kang2019numerical,sun2020numerical}, which are based on three-dimensional computations.}

Because $\Lew_O$ is taken to be unity, a simplified formulation of the above system of equations can be obtained~\cite{linan1991structure} by introducing the mixture fraction $Z$ and the excess enthalpy variable $H$,
\begin{equation}
    Z = \frac{SY_F-Y_O+1}{S+1}, \qquad H = Y_F+Y_O-1+\theta, \label{Z H}
\end{equation}
which are normalized such that $Z-1=H=0$ at $y=-1$ and $Z=H=0$ at $y=+1$. A linear combination of~\eqref{yF nondim}-\eqref{theta nondim} yields the chemistry free equation
\begin{equation}
    \left[\pfr{}{t} - (1+p^2)\sd{}{x} - \sd{}{y}\right][SH-(S+1)Z]=0,  \nonumber
\end{equation}
 which upon integration leads to
\begin{equation}
    Z - Z_0 = \frac{SH}{S+1}, \quad \text{where} \quad Z_0 \equiv \frac{1}{2}(1-y). \label{Z0}
\end{equation}

Consequently the problem can be described in terms of two dependent variables only, which are selected here to be $H$ and $Y_F$. These are governed by
\begin{align}
      \pfr{H}{t} - \left(\sd{H}{X} + \sd{H}{y}\right)  &=  -\left[\frac{(\Lew_x-1)}{\Lew_x}\sd{Y_F}{X} + \frac{(\Lew-1)}{\Lew}\sd{Y_F}{y} \right] \label{H final}\\
   \pfr{Y_F}{t} - \left(\frac{1}{\Lew_x}\sd{Y_F}{X} + \frac{1}{\Lew}\sd{Y_F}{y}\right) &= -\Dam\, \omega (H,Y_F), \label{YF final}
\end{align}
where the scaled variable $X=x/\sqrt{1+p^2}$ is introduced for convenience and where $\omega(H,Y_F)$ is obtained by substituting the equations
\begin{align}
    Y_O &= 1 - S(H-Y_F)  - (S+1)Z_0 \label{YO final}\\
    \theta &= (S+1)(H-Y_F+Z_0) , \label{theta final}
\end{align}
which follow from~\eqref{Z H} and~\eqref{Z0} into the expression of $\omega$ given by~\eqref{omega}. It is worth noting in~\eqref{H final}-\eqref{YF final} the presence of the longitudinal Lewis number $\Lew_x$ defined in~\eqref{Lex}. When $\Lew=1$, we have $\Lew_x=1$ and $H=0$ and one needs to solve only for $Y_F$.

\section{The linear stability problem for the planar diffusion flame}

\subsection{The base solution}
The base solution whose stability is under investigation corresponds to the steady planar diffusion flame aligned with the flow direction. Since this solution, denoted by an overbar, depends only on the $y$ coordinate, it is governed by~\eqref{H final}-\eqref{YF final} with $\pfi{}{t}=\pfi{}{X}=0$, that is by equations
\begin{align}
    \frac{d^2\overbar{H}}{dy^2}  &=  \frac{\Lew-1}{\Lew} \frac{d^2\overbar{Y}_F}{dy^2}  \nonumber\\
    \frac{1}{\Lew}\frac{d^2\overbar{Y}_F}{dy^2} &= \Dam\, \omega(\overbar{H},\overbar{Y}_F),  \nonumber
\end{align}
subject to the boundary conditions $\overbar{H}(-1)=\overbar{Y}_F(-1)-1=\overbar{H}(+1)=\overbar{Y}_F(+1)=0$. The first equation is integrated to obtain
\begin{align}
      \overbar{H} =\frac{(\Lew-1)}{\Lew} (\overbar{Y}_F-Z_0)   \nonumber
\end{align}
whereas the second equation is integrated numerically. \blue{Specifically, the nonlinear differential equation is solved numerically using COMSOL Muliphysics software, which determines $\Dam$ as an unknown parameter along with the solution for values of the temperature prescribed at a given location, sat at $y=0$.}

\begin{figure}[h!]
\centering
\includegraphics[scale=0.7]{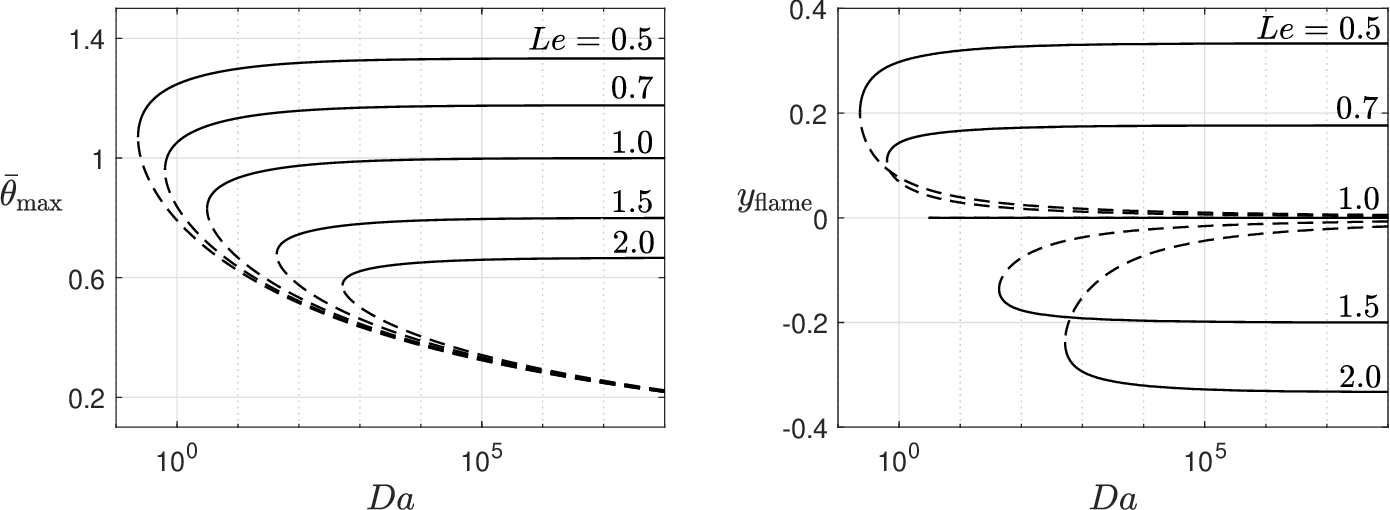}
\caption{Maximum temperature and flame location of the unperturbed planar diffusion flame as functions of $\Dam$ for selected values of $\Lew$; here and elsewhere $\alpha=0.85$, $\beta=10$ and $S=1$.} \label{fig:Scurve}
\end{figure}

A sample of the numerical results is given in Fig.~\ref{fig:Scurve}. Shown are plots of the maximum of the temperature field, $\overbar\theta_{\mathrm{max}}$, and  its location, $y_{\mathrm{flame}}$, versus the Damk\"{o}hler number $\Dam$ for selected values of $\Lew$. In the limit $\Dam\rightarrow \infty$, these quantities are seen to approach their Burke-Schumann values
\begin{equation}
    y_{\mathrm{flame}} = \frac{S-\Lew}{S+\Lew}, \qquad \overbar\theta_{\mathrm{max}} = \frac{S+1}{S+\Lew}.   \nonumber
\end{equation}
The solid curves correspond to near-equilibrium diffusion flames and the dashed curves to flames in the so-called partial burning regime~\cite{linan1974asymptotic} which are typically unstable. For each $\Lew$, the turning point $\Dam=\Dam_{\rm{ext}}$ corresponds to extinction. Since $\Dam_{\rm{ext}}$ varies significantly with $\Lew$ as evident from the figure, it is convenient for later reference to use
\begin{equation}
    \delta = \frac{\Dam-\Dam_{\rm{ext}}}{\Dam_{\rm{ext}}}  \label{delta}
\end{equation}
as the relative measure of the Damk\"{o}hler number. 

\subsection{Perturbed state}
To investigate the stability of the planar diffusion flame, we consider infinitesimal normal-mode perturbations superimposed on the base solution such that
\begin{align}
    Y_F &= \overbar{Y}_F(y) + \widetilde{Y}_F(y) e^{i\kappa X + \sigma t}, \qquad |\widetilde{Y}_F|\ll |\overbar{Y}_F|  \nonumber\\
    H &= \overbar{H}(y) + \widetilde{H}(y) e^{i\kappa X + \sigma t}, \qquad |\widetilde{H}|\ll |\overbar{H}| \nonumber
\end{align}
where $\sigma $ is the complex growth rate and $\kappa $ is the real wave number. Similarly, one may define the perturbations $\widetilde{Y}_O(y)$ and $\widetilde{\theta}(y)$, which according to~\eqref{YO final}-\eqref{theta final} are related to $\widetilde{Y}_F(y)$ and $\widetilde{H}(y)$ by $\widetilde{Y}_O = S(\widetilde{Y}_F -  \widetilde{H})$ and $\widetilde{\theta} = (S+1)(\widetilde{H} -  \widetilde{Y}_F)$.

Substituting the perturbed dependent variables in~\eqref{H final}-\eqref{YF final} and linearizing, we obtain the eigenvalue problem
\begin{align}
     \left(\frac{d^2}{dy^2} - \kappa^2\right) \widetilde{H} -  \left[\frac{(\Lew-1)}{\Lew}\frac{d^2}{dy^2} -\frac{(\Lew_x-1)\kappa^2}{\Lew_x}\right]\widetilde{Y}_F &=\sigma \widetilde{H} \label{H eig}\\
    \left(\frac{1}{\Lew} \frac{d^2}{dy^2} -  \frac{\kappa^2}{\Lew_x}-f\right) \widetilde{Y}_F -g \widetilde{H} &= \sigma\widetilde{Y}_F, \label{YF eig}
\end{align}
where
\begin{align}
    f(y) &= \beta^3 \Dam \left[S\overbar Y_F + \overbar Y_O -\frac{\beta (S+1) \overbar Y_F\overbar Y_O }{[1-\alpha(1-\overbar\theta)]^2} \right] \exp\left[\frac{\beta(\overbar\theta-1)}{1+\alpha(\overbar\theta-1)}\right],  \nonumber\\
    g(y) &= \beta^3 \Dam \left[-S\overbar Y_F+\frac{\beta (S+1) \overbar Y_F\overbar Y_O }{[1-\alpha(1-\overbar\theta)]^2} \right] \exp\left[\frac{\beta(\overbar\theta-1)}{1+\alpha(\overbar\theta-1)}\right].  \nonumber
\end{align}
The quantities $\widetilde{H}$ and $\widetilde{Y}_F$ satisfy homogeneous Dirichlet boundary conditions, namely,
\begin{equation}
    \widetilde{H}=0, \quad  \widetilde{Y}_F=0 \quad \text{at} \quad y=\pm 1. \label{BC eig}
\end{equation}
This eigenvalue problem possesses a discrete spectrum of eigenvalues $\sigma_n$ indexed by an integer $n$. For example, when $\delta\rightarrow \infty$, the eigenvalues are given by a simple formula, derived in Appendix A. In general, the eigenvalue with the largest real part (real growth rate), $\max\limits_n[\mathrm{Re}\{\sigma_n\}]$, determines the stability characteristics. Henceforth, $\sigma$ shall refer to this eigenvalue and not to the entire spectrum.

\section{Flame stability and bifurcation diagrams}
\label{disp}
\subsection{Preliminary remarks and notations}
\label{prelim}
The main object of our stability analysis is to describe the relationship
\begin{equation}
    \sigma=\sigma(\kappa;\Lew,p,\delta) \label{sigma}
\end{equation}
 between $\sigma$ (the eigenvalue with the largest real part), the wavenumber $\kappa$ and the parameters $\Lew$, $p$ and $\delta$. All eigenvalues are computed by solving the eigenvalue problem~\eqref{H eig}-\eqref{BC eig} numerically \blue{with a tolerance of $10^{-6}$} using the Chebfun package~\cite{driscoll2014chebfun}, \blue{which is based on a spectral collocation method using Chebyshev polynomials.}

 In doing so, computations are restricted to the strongly burning, near-equilibrium diffusion flames, the top branches in the left plot of Fig.~\ref{fig:Scurve}. As the scaled Damk\"{o}hler number $\delta$ defined in~\eqref{delta} tends to infinity, the corresponding Burke-Schumann flame is known to be unconditionally stable when $p=0$~\cite{kim1996diffusional}. This conclusion remains true for $p\neq 0$ as shown in Appendix A. As $\delta$ is decreased the flame becomes unstable for values of $\delta$ smaller than a critical value $\delta_c$, determined by the marginal stability condition $\mathrm{Re}\{\sigma\}=0$. Planar adiabatic near-equilibrium diffusion flames are thus unstable in the range $0<\delta<\delta_c$.

The flame stability for $p=0$ has been investigated in~\cite{kim1996diffusional} and~\cite{kim1997linear} through asymptotic analyses in the so-called slowly varying flame and near-equidiffusional flame limits~\cite{buckmaster1982theory}, respectively. In particular, it has been shown in~\cite{kim1997linear,vance2001stability} that diffusion flames are unstable for $\Lew>\Lew_c$, where $\Lew_c>1$ is a critical value depending on $\delta$. The ensuing instability was found to be a longwave instability with or without pulsations in time.

In our problem allowing for non-zero values of $p$, we also find that no instability occurs for $\Lew<\Lew_c$ where $\Lew_c=\Lew_c(\delta,p)>1$. However, unlike in the $p=0$ case, the instability need not always be a longwave instability, i.e., an instability for which the most unstable mode corresponds to $\kappa=0$. Indeed, we also find for a certain parametric range, finite wavelength instability for which the most unstable mode corresponds to  $\kappa=\kappa_m\neq 0$, as we shall confirm below.

To facilitate the discussion, it is convenient to introduce the notation
\begin{equation}
    \sigma_0\equiv \sigma(\kappa=0) \qquad  \text{and} \qquad \sigma_m \equiv \sigma(\kappa=\kappa_m) \quad \text{with} \quad \kappa_m\neq 0,   \label{sig0m}
\end{equation}
where the values $\sigma_0$ and $\sigma_m$ correspond to the condition $d\sigma/d\kappa=0$\footnote{Since the function $\sigma(\kappa)$ involves $\kappa^2$ only (see~\eqref{H eig}-\eqref{YF eig}), it is an even function of $\kappa$ and thus $d\sigma/d\kappa$ is always zero at $\kappa=0$. Therefore, $\sigma_0$ always exists and corresponds either to an extremum or a saddle point. Furthermore, as $\kappa\rightarrow 0$, $\Lew_x$ and consequently $p$ disappear from equations~\eqref{H eig}-\eqref{YF eig}. Therefore $\sigma_0$ depends on $\Lew$ and $\delta$, but not on $p$. In contrast, the existence of $\sigma_m$ is not always warranted and depends on $p$.}. In fact, we may regard $\sigma_0$ as the growth rate (possibly complex) characterising the development of the longwave instability. In this case, the instability will be appropriately referred to as a \textit{non-oscillatory longwave instability} if $\sigma_0$ is real and as an \textit{oscillatory longwave instability} otherwise. Similarly, $\sigma_m$ may be regarded as the growth rate (found always to be real) characterising the development of the \textit{finite wavelength (cellular) instability}.

\subsection{Results for cases with $\delta$ varying and fixed values of $p$ and $\Lew$}

\begin{figure}[h!]
\centering
\includegraphics[scale=0.7]{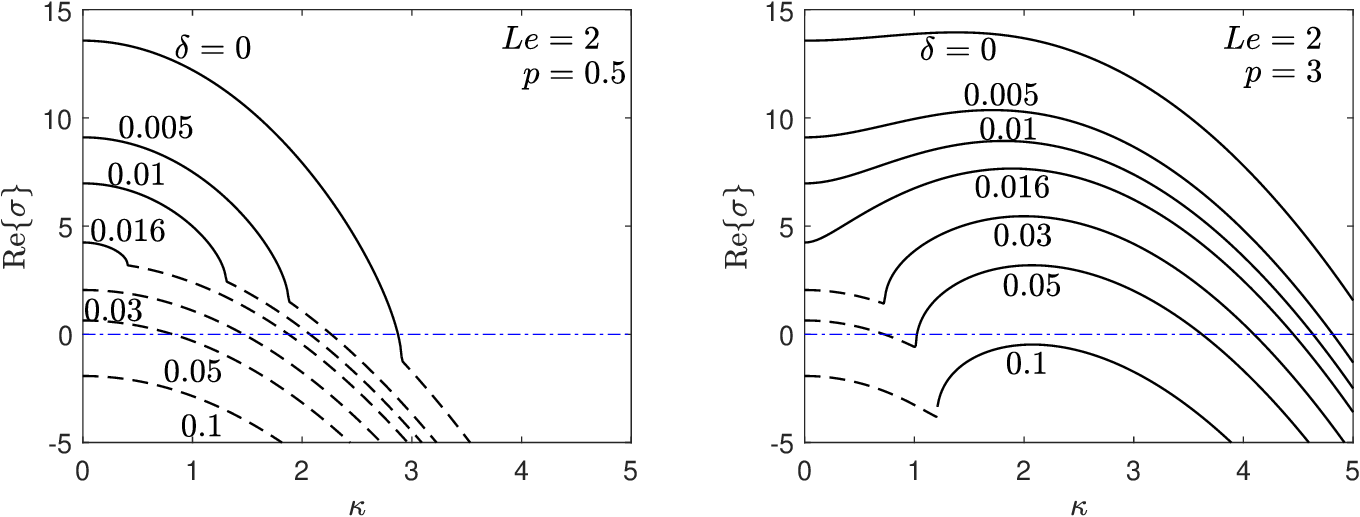}
\caption{ Maximum growth rate, $\mathrm{Re}\{\sigma\}$, versus the wave number $\kappa$ for selected values of $\delta$. The left plot is calculated for $\Lew=2$ and $p=0.5$ and  the right plot for $\Lew=2$ and $p=0.5$. Solid lines represent   real branches 
(indicating that $\mathrm{Im}\{\sigma\}=0$) and   dashed lines  complex branches ($\mathrm{Im}\{\sigma\} \neq 0$). All computations are carried out with $\beta=10$, $\alpha=0.85$ and $S=1$.} \label{fig:Le2}
\end{figure}

The effect of varying $\delta$ on flame stability \blue{for the top-branches in the left subfigure of Fig.~\ref{fig:Scurve}} is illustrated in Fig.~\ref{fig:Le2} pertaining to $\Lew=2$ and $p=0.5$ (left subfigure) and $p=3$ (right subfigure). Plotted is the real part of $\sigma$ versus the wave number $\kappa$ for selected values of $\delta$. The corresponding curves (dispersion curves) are drawn with solid lines when $\sigma(\kappa)$ is real and dashed lines when $\sigma(\kappa)$ has a non-zero imaginary part.

The left subfigure corresponding to $p=0.5$ is found to be qualitatively similar to the $p=0$ case (not shown); here the most unstable mode always corresponds to $\kappa=0$, indicating the presence of a longwave instability. This longwave instability is non-oscillatory when $0<\delta<\delta_\ast$ and oscillatory when $\delta_\ast<\delta<\delta_c$, where $\delta_\ast$ is some critical value ($\delta_\ast \approx 0.0168$ when $\Lew=2$ irrespective of the value of $p$; see below).

We turn now to the right subfigure which corresponds to $p=3$. Here the notable feature is  that the most unstable mode  corresponds to a non-zero value $\kappa_m$ of $\kappa$. Therefore we are in the presence of a finite wavelength (cellular) instability. Furthermore, it is worth pointing out that the dispersion curve is associated with non-oscillatory modes characterized by real growth rates for $0<\delta<\delta_\ast$ (solid lines in the subfigure), but that unstable oscillatory modes also exist when $\delta_\ast<\delta<\delta_c$ (dashed lines).

\begin{figure}[h!]
\centering
\includegraphics[scale=0.7]{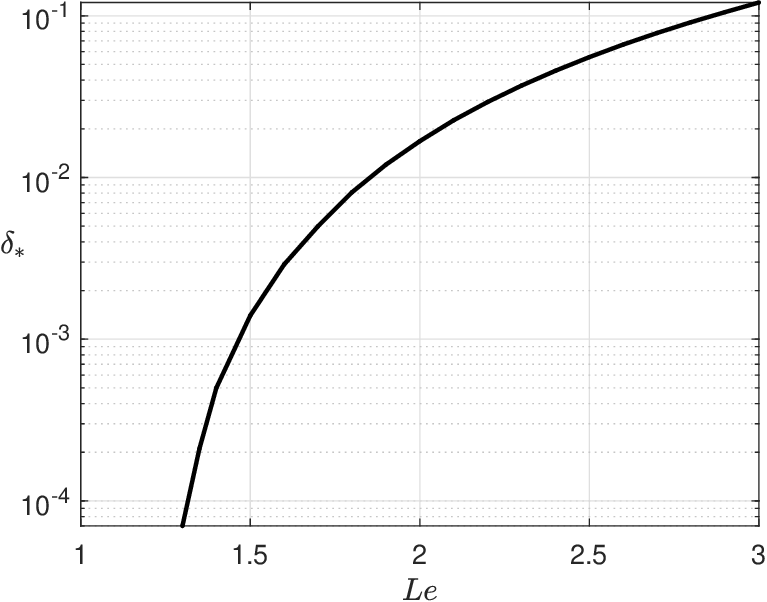}
\caption{The parameter $\delta_\ast$ versus $\Lew$, computed for $\beta=10$, $\alpha=0.85$ and $S=1$. This parameter characterizes the transition from  a non-oscillatory longwave instability (obtained when $0<\delta<\delta_\ast$) to an oscillatory longwave instability (obtained when $\delta_\ast<\delta<\delta_c$).} \label{fig:deltas}
\end{figure}

Before closing this section, we note that the parameter $\delta_\ast$ introduced above corresponds to $\sigma_0$ defined in~\eqref{sig0m} changing, as $\delta$ is increased, from being a real number to being a complex number with nonzero imaginary part. Since $\sigma_0$ is independent of $p$ (see footnote 1), $\delta_\ast$, a function of $\Lew$, is also independent of $p$. This function is plotted in Fig.~\ref{fig:deltas} and the curve thus computed characterizes the transition from oscillatory to non-oscillatory longwave instabilities.

\subsection{Results for cases with $p$ varying and fixed values of $\delta$ and $\Lew$}
\label{jc}
\begin{figure}[h!]
\centering
\includegraphics[scale=0.7]{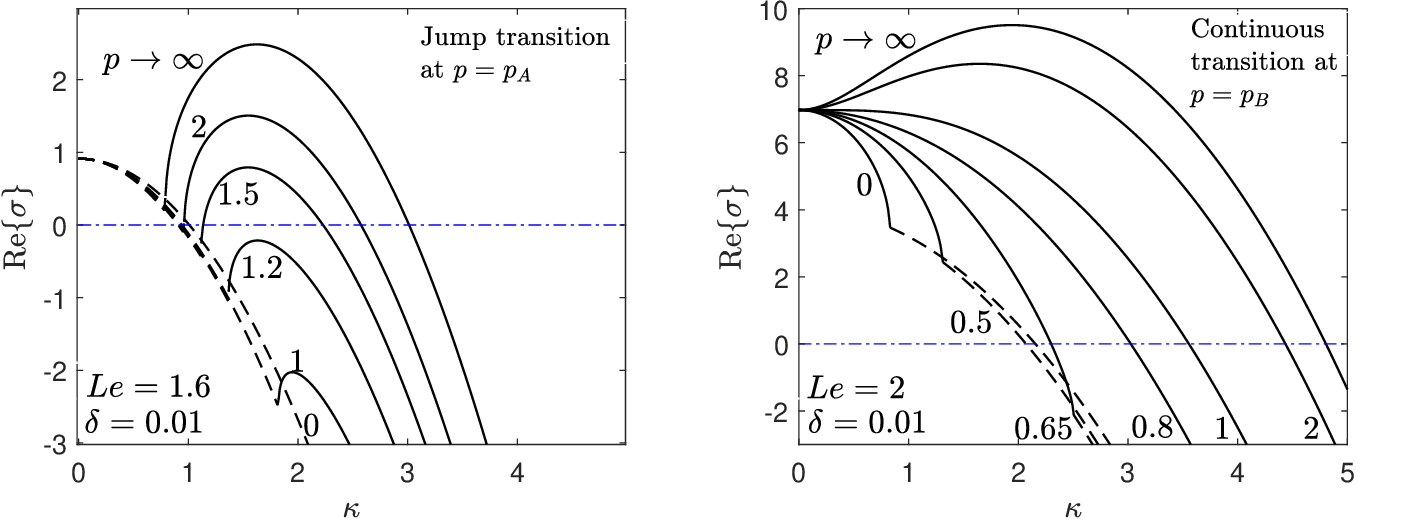}
\caption{Dispersion curves  representing the maximum growth rate, $\mathrm{Re}\{\sigma\}$, versus the wave number $\kappa$ for selected values of $p$.  The left subfigure summarizes calculations for $\Lew=1.6$ and $\delta=0.01$ and the right subfigure for $\Lew=2$ and $\delta=0.01$. All computations are carried out with $\beta=10$, $\alpha=0.85$ and $S=1$. Solid lines represent real branches ($\mathrm{Re}\{\sigma\}=\sigma$) and dashed lines complex branches ($\mathrm{Im}\{\sigma\}\neq0$). The values of critical Peclet numbers are $p_A\approx 1.56$ and $p_B\approx 1.00$.} \label{fig:delta0.01}
\end{figure}

In this section, we investigate  the effects of varying $p$ for a fixed value of $\delta$. Variations in $p$ can in fact be achieved conceptually by adjusting the flow rate as done in experiments~\cite{miesse2005experimental,miesse2005diffusion}. An illustrative case is shown in Figure~\ref{fig:delta0.01} corresponding to $\delta=0.01$ and $\Lew=1.6$ (left subfigure) and $\Lew=2$ (right subfigure). Plotted is the real part of $\sigma$ versus the wave number $\kappa$ for selected values of $p$. As in figure~\ref{fig:Le2}, the sections of the dispersion curves drawn with solid lines correspond to $\sigma(\kappa)$ being real and dashed sections to $\sigma(\kappa)$ having a non-zero imaginary part. It can be seen that $\sigma_0\equiv\sigma(\kappa=0)$ (as defined in~\eqref{sig0m}) is independent of $p$, in line with the remark made in footnote 1.

Focusing first on the left subfigure, pertaining to $\Lew=1.6$, we  note that the dispersion curve for $p=0$ is entirely complex (plotted as a dashed line) and as its maximum occurs at $\kappa=0$, it indicates  the presence of an oscillatory longwave instability. On the other hand, for $p=1$, the dispersion curve consists of a complex branch (dashed line) which meets a real branch (solid line) at a finite value of $\kappa$ where a cusp is encountered.  Although for this value of $p$, the real branch is still in the stable region, it gradually crawls up on the complex branch as $p$ is increased, thus narrowing the range of oscillatory modes. The maximum of the real branch $(\kappa_m,\sigma_m)$ enters the unstable region as $p$ exceeds a certain value of $p$ corresponding to $\sigma_m=0$. When $p$ is increased further, a critical value $p=p_A$ is encountered for which $\sigma_m=\mathrm{Re}\{\sigma_0\}$. For $p>p_A$, $\sigma_m>\mathrm{Re}\{\sigma_0\}$ and therefore the most unstable mode occurs at $\kappa=\kappa_m\neq 0$. This indicates the presence of a finite wavelength (cellular) instability. The transition between the oscillatory longwave instability (occurring for $p<p_A$)  and the finite wavelength instability (occurring for $p>p_A$) may be viewed as a jump transition, in the sense that a cellular pattern is predicted to suddenly emerge as soon as $p$ exceeds $p_A$.

We turn now to the right subfigure to examine how the findings above are affected by adopting a higher value of the Lewis number, $\Lew=2$. Here, the dispersion curve for $p=0$ implies a non-oscillatory longwave instability, unlike the corresponding curve in the left subfigure where an oscillatory longwave instability occurs when $p=0$. As can be seen, an increase in $p$ leads the concavity of the dispersion curve at $\kappa=0$ to change from being negative to positive as $p$ crosses a critical value $p=p_B$, corresponding to the condition $d^2\sigma/d\kappa^2|_{\kappa=0}=0$. For $p>p_B$, the dispersion curve has a maximum at $\kappa_m\neq 0$, implying the presence of a finite wavelength instability. For $p<p_B$, the dispersion curve peaks at  $\kappa=0$ and no maximum occurs at a finite values of $\kappa$ and the instability is a non-oscillatory longwave one. Note however that the transition at $p=p_B$ between the non-oscillatory longwave instability and the finite wavelength instability is continuous in the sense that $\kappa_m\to 0$ as $p$ decreases towards $p_B$.

\subsection{Stability regime diagram}
\label{regime}
\begin{figure}[h!]
\centering
\includegraphics[scale=0.75]{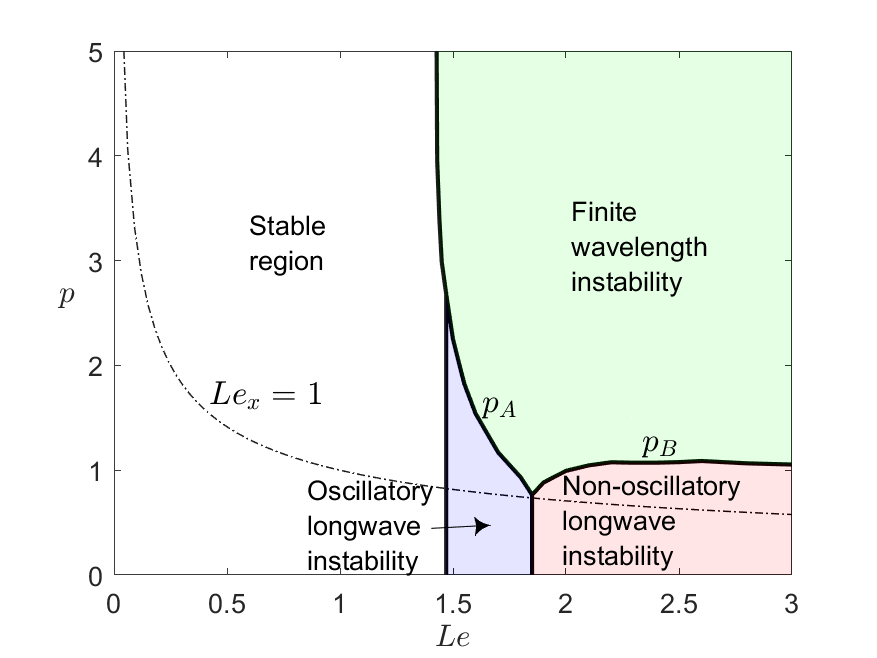}
\caption{Stability regime diagram of a planar diffusion flame aligned with the direction of a shear flow, computed for $\delta=0.01$, $\alpha=0.85$, $\beta=10$ and $S=1$.} \label{fig:regime}
\end{figure}

Based on our discussion above of the various types of instabilities and transitions, we can synthesize the findings in the form of a regime diagram in the $\Lew$-$p$ plane.  This is computed and shown in Fig.~\ref{fig:regime} in the illustrative case $\delta=0.01$. Solid lines in this diagram separate regions possessing different stability characteristics, whereas the dash-dotted curve represents the equation $p=1/\sqrt{\Lew}$ which is equivalent, according to~\eqref{Lex asym}, to the condition $\Lew_x=1$. We note that in the domain above this dash-dotted curve, we have $\Lew_x<1$ if $\Lew>1$ and $\Lew_x>1$ if $\Lew<1$. Four regions can be identified in the diagram: a stable region on the left (white region) and three unstable regions on the right (shaded regions), corresponding to  the three types of instabilities encountered.

The boundary curve separating the stable region from the unstable ones translates the critical condition $\Lew=\Lew_c(\delta,p)$ for $\delta=0.01$ with $\Lew_c$ being as introduced in $\S$\ref{prelim}. This boundary curve consists of two sections. The lower section, consisting of the vertical line $\Lew_c\approx 1.47$ for $p\in[0,2.68]$, corresponds to a bifurcation associated with an oscillatory longwave instability; this is often referred to in the literature as a type-III$_{o}$ instability~\cite{cross1993pattern}. The upper section, occurring for $p\in[2.68,\infty)$ and $\Lew_c\in[1.43,1.47]$, exhibits a weak dependence of $\Lew_c$ on $p$. This section corresponds to a bifurcation associated with a non-oscillatory (or stationary) finite wavelength instability, often referred to as a type-I$_{s}$ instability~\cite{cross1993pattern}.

We now briefly comment on the boundaries between the three unstable regions aforementioned, which are seen to intersect at a cusp point $(\Lew,p)=(1.85,0.76)$. First, the vertical line at $\Lew\approx 1.85$ for $p\in[0,0.76]$ is obtained from Fig.~\ref{fig:deltas} using the condition $\delta_*=0.01$. The curve, labeled $p_A$, corresponding to the critical condition $p=p_A$ introduced in the previous subsection, separates the regions of oscillatory longwave instability and finite wavelength instability. Similarly, the curve, labeled $p_B$, corresponds to the critical condition $p=p_B$ and separates the regions of non-oscillatory longwave instability and finite wavelength instability. An important observation related to this figure is that the finite wavelength instability region lies entirely in the region $\Lew_x<1$ (above the dash-dotted curve). This demonstrates that a necessary condition for the occurrence of the finite wavelength (or cellular) instability is that $\Lew_x<1$ provided the fuel Lewis number $\Lew>1$.

\begin{figure}[h!]
\centering
\includegraphics[scale=0.75]{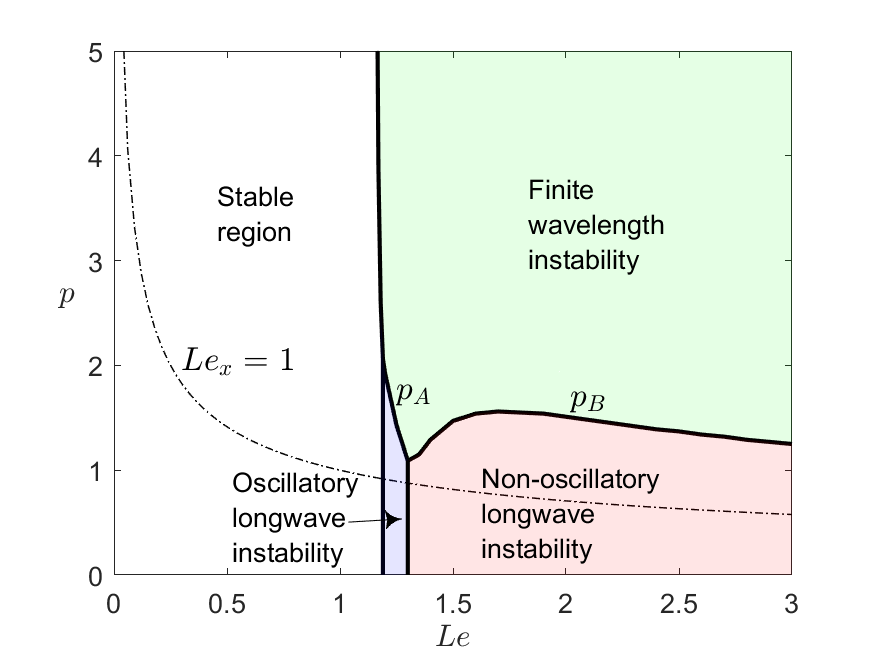}
\caption{Stability regime diagram of a planar diffusion flame aligned with the direction of a shear flow, computed for $\delta=0.0001$, $\alpha=0.85$, $\beta=10$ and $S=1$.} \label{fig:regime2}
\end{figure}

\blue{It is also interesting to} examine how a change in $\delta$ affects the stability regime diagram of Fig.~\ref{fig:regime}, by carrying out similar computations for $\delta=10^{-4}$. The results are summarized in Fig.~\ref{fig:regime2}. The figure shows that the four regions identified earlier persist in this case, except that the unstable regions are extended now further to the left ($\Lew_c \approx 1.17$), thereby narrowing the stable region. Moreover, the extent of the oscillatory longwave instability region is found now to be narrower in comparison with that of Fig.~\ref{fig:regime}. 

\blue{To close this section, we note that our results have been computed for the stoichiometrically balanced case, $S=1$. However,  it is possible to provide qualitative comments relevant to other values of $S$, since the controlling Lewis number pertains to the deficient reactant which corresponds to the fuel when $S\ll 1$ and to the oxidizer when $S\gg 1$. Furthermore, since the influence of Taylor dispersion on the flame instability is stronger when the Lewis number is further from unity, this influence is more apparent when $S\ll 1$ than when $S\gg 1$ because $\Lew_O\approx 1$. In particular, we can infer that cellular instabilities are easier to realize in mixtures with $\Lew_F=\Lew>1$ and $S\ll 1$ and more difficult in mixtures with $\Lew_F=\Lew<1$ and $S\ll 1$, a fact which is consistent with experiments~\cite{miesse2005diffusion}.}

\section{Time dependent numerical simulations}

The stability regime diagram of $\S$\ref{regime}, based on the computations of the eigenvalues of the linear stability problem~\eqref{H eig}-\eqref{BC eig}, summarizes the main conclusions that one could infer from the linear analysis. To describe the eventual fate of infinitesimal disturbances, we need to solve the full nonlinear problem~\eqref{H final}-\eqref{YF final}. This will allow, in particular, to answer the important question about the existence of cellular structures and their stability. The solution of equations~\eqref{H final}-\eqref{YF final} is carried out numerically and selected cases in each of the three unstable regions of diagram~\ref{fig:regime} are presented herein. The computations are performed using COMSOL Multiphysics, a package which has been extensively tested in combustion applications, see e.g.~\cite{pearce2014taylor,daou2018taylor,daou2021effect,rajamanickam2023thick,daou2023flame}. The computational domain, taken to be $(-15,15)\times(-1,1)$, is covered by a grid consisting of approximately 400,000 triangular elements and includes local refinement in the reaction zone. Solutions have been tested to be independent of the mesh and of the time step $\Delta t$ (whose maximum value is set to be $\Delta t=0.005$). Dirichlet boundary conditions corresponding to~\eqref{nondim BC1} and~\eqref{nondim BC2} are applied in the $y$-direction. As far as the $X$-direction, we adopt periodic boundary conditions. These boundary conditions are suitable for our idealized simple model, given that our main concern is to study the stability of a planar diffusion flame aligned with the direction of the flow. In particular, we disregard complicating effects associated with heat loss and with special inlet/outlet conditions in finite domains which are encountered in practice.

\subsection{Longwave instability}

\begin{figure}[h!]
\centering
\includegraphics[scale=0.7]{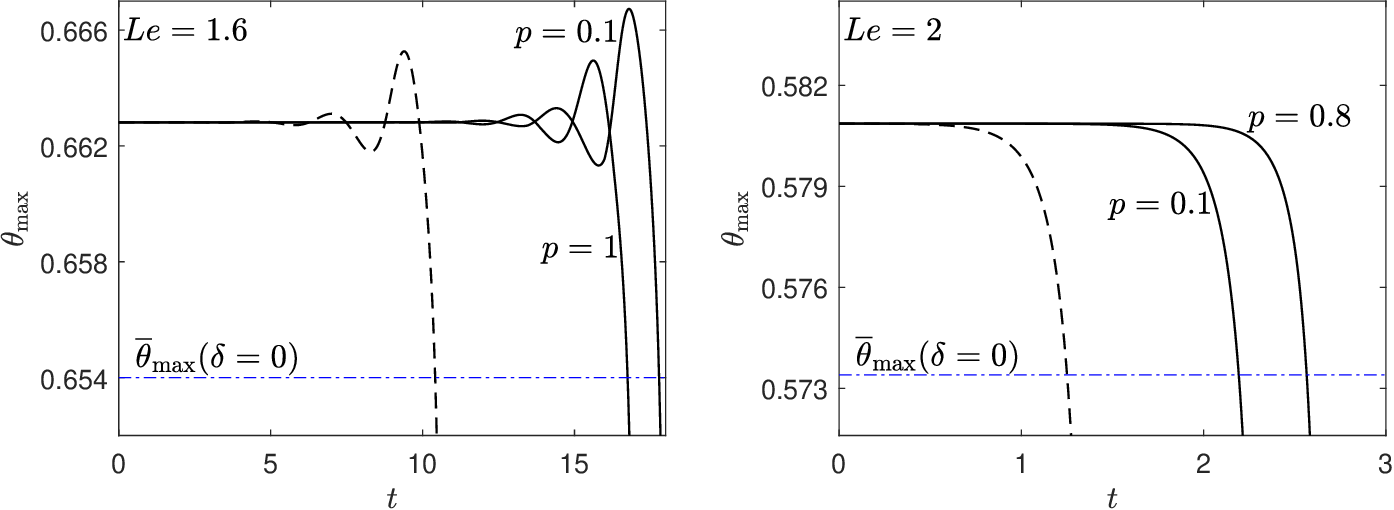}
\caption{Maximum temperature ${\theta}_{\mathrm{max}}$ as a function of time $t$, calculated for $\Lew=1.6$ (left) and $\Lew=2$ (right) with $\delta=0.01$. The dashed lines pertain to one-dimensional computations (independent of $p$), whereas the solid lines pertain to two-dimensional computations for selected values of $p$. All computations are carried out with $\beta=10$, $\alpha=0.85$ and $S=1$.} \label{fig:longwave}
\end{figure}

The two types of longwave instability, namely oscillatory and non-oscillatory, identified through our linear stability analysis, are always found in our computations to ultimately result in flame quenching\footnote{In this paper we are not concerned with the problem of stability to finite-amplitude perturbations. The dynamics of the diffusion flame under such perturbations is studied in~\cite{sohn1999instability,sohn2000nonlinear,sung2002stretch}. These studies show that finite-amplitude perturbations can lead a linearly stable flame either into a permanently oscillating state (a limit cycle) or to a quenched state depending on the amplitude of the perturbation.}. The only difference between these two types is found to lie in their transient evolution to extinction. This is illustrated in Fig.~\ref{fig:longwave} for two cases corresponding to $\Lew=1.6$ (left subfigure) and $\Lew=2$ (right subfigure) with selected values of $p$. Shown in this figure is the maximum temperature $\theta_{\mathrm{max}}$ versus time $t$. We note that the solid curves correspond to two-dimensional computations, while the dashed curves, pertaining to $\kappa=0$, are computed as solutions of the one-dimensional problem obtained from~\eqref{H final}-\eqref{YF final} by dropping the $X$-derivatives. The two-dimensional computations are carried out with the horizontal domain size equal to $30$, as mentioned above, and therefore the minimum wavenumber allowed numerically is $\kappa=2\pi/30=0.2094$. We note that once $\theta_{\mathrm{max}}(t)$ drops below the steady-state temperature $\overbar{\theta}_{\mathrm{max}}$ corresponding to the static extinction condition $\delta=0$ (the turning points in Fig.~\ref{fig:Scurve}), ${\theta}_{\mathrm{max}}$ cannot recover and decays monotonically to zero.

\begin{figure}[h!]
%\centering
\advance\leftskip-2cm
\includegraphics[scale=0.75]{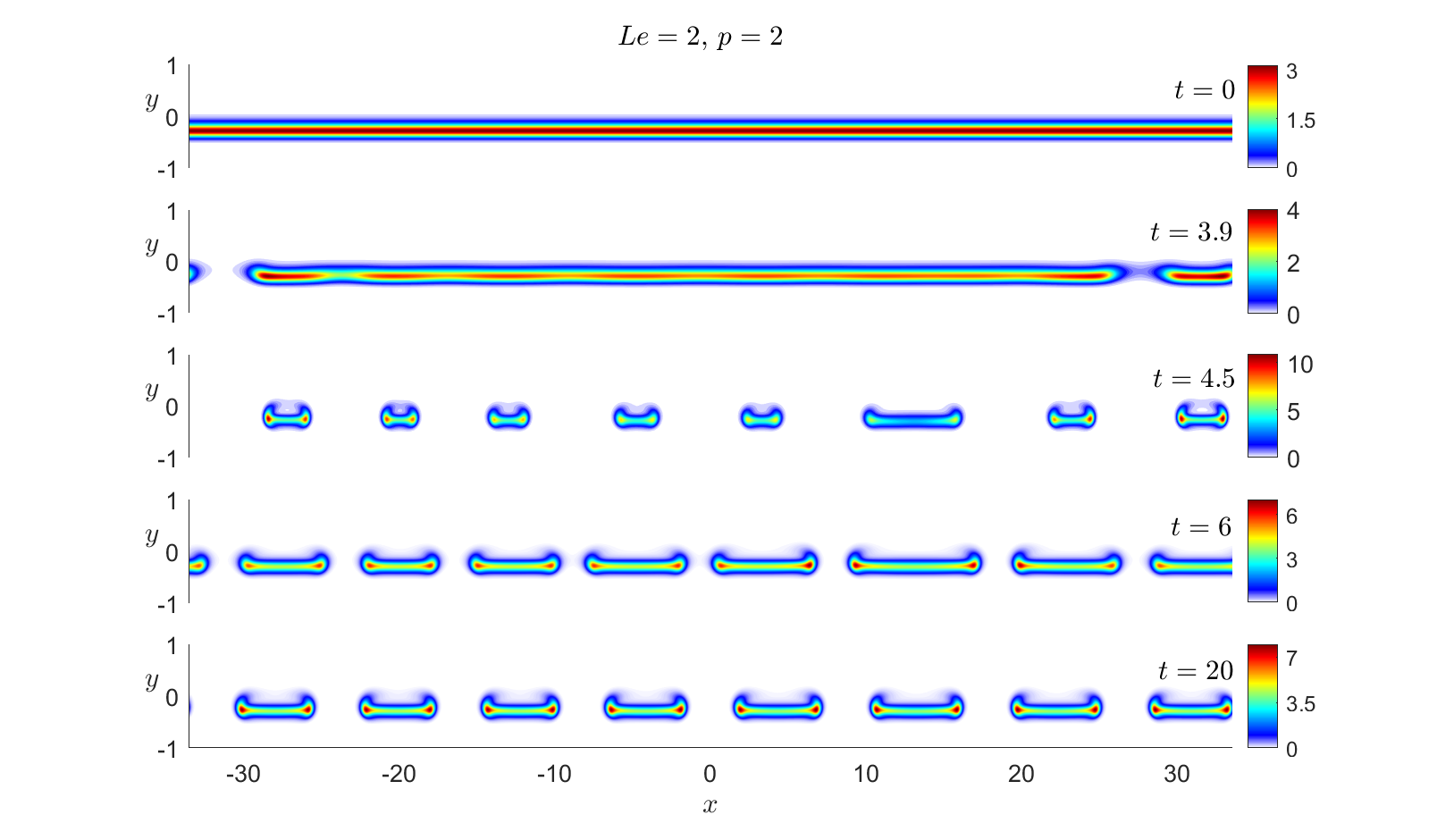}
\vspace{-1cm}
\caption{Reaction rate ($\omega$) fields at selected values of time $t$ for $\Lew=2$, $p=2$, $\delta=0.01$, $\beta=10$, $\alpha=0.85$ and $S=1$.} \label{fig:Le2p2time}
\end{figure}

\subsection{Finite wavelength instability}

\begin{figure}[h!]
%\centering
\advance\leftskip-2cm
\includegraphics[scale=0.75]{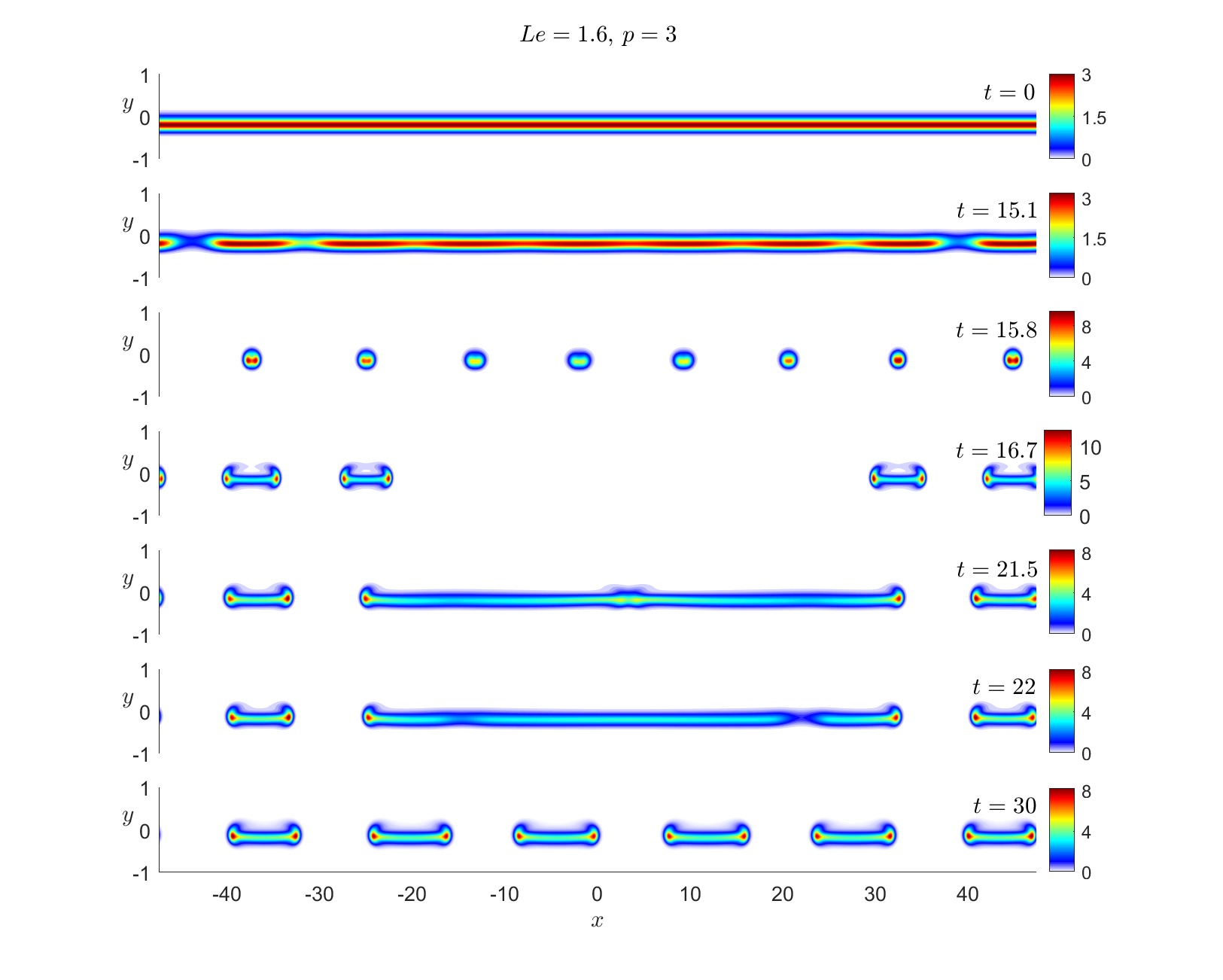}
\vspace{-1cm}
\caption{Reaction rate ($\omega$) fields at selected values of time $t$ for $\Lew=1.6$, $p=3$, $\delta=0.01$, $\beta=10$, $\alpha=0.85$ and $S=1$.} \label{fig:Le1.6p3time}
\end{figure}

\begin{figure}[h!]
%\centering
\advance\leftskip-2cm
\includegraphics[scale=0.75]{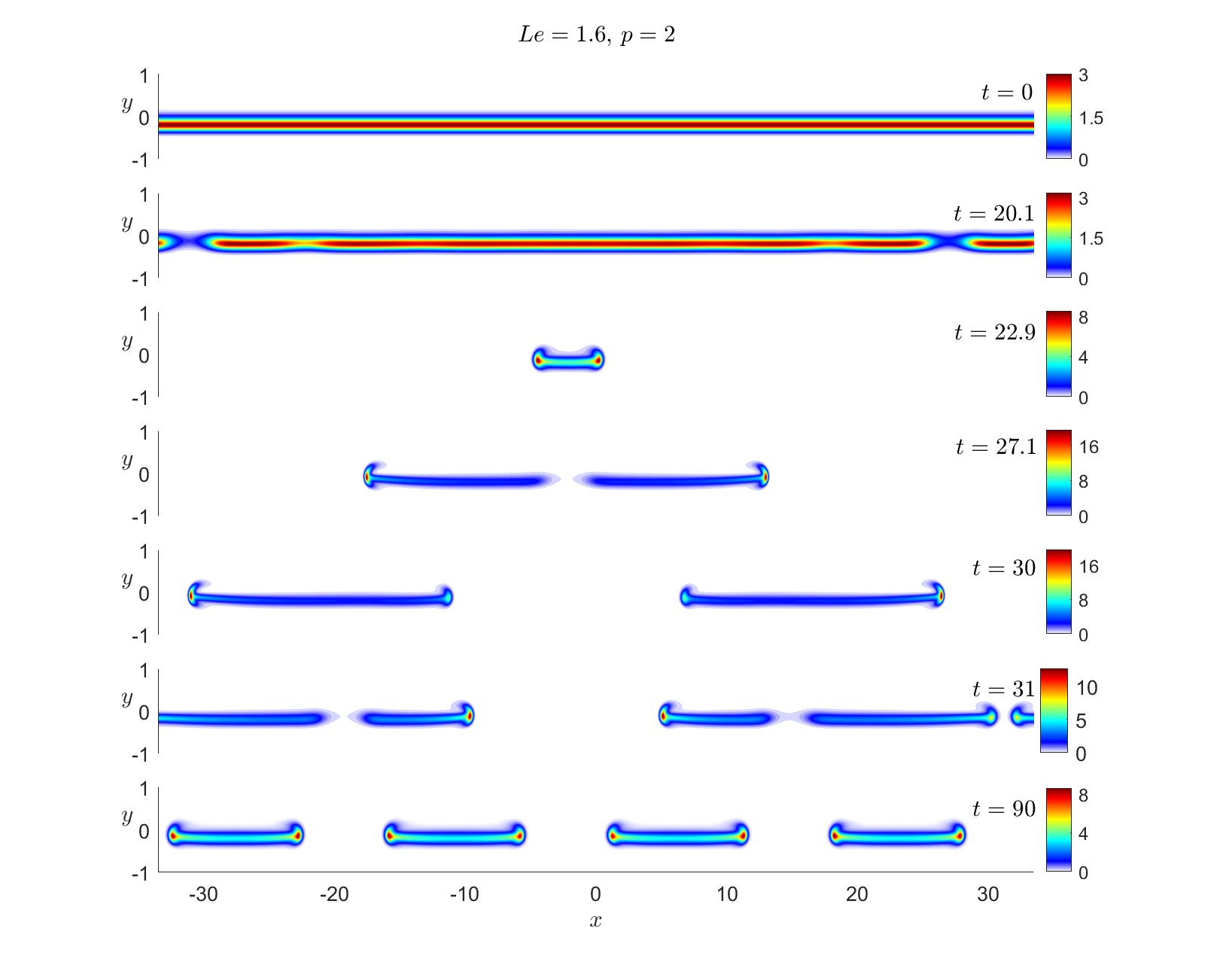}
\vspace{-1cm}
\caption{Reaction rate ($\omega$) fields at selected values of time $t$ for $\Lew=1.6$, $p=2$, $\delta=0.01$, $\beta=10$, $\alpha=0.85$ and $S=1$.} \label{fig:Le1.6p2time}
\end{figure}

\begin{figure}[h!]
%\centering
\advance\leftskip-2cm
\includegraphics[scale=0.75]{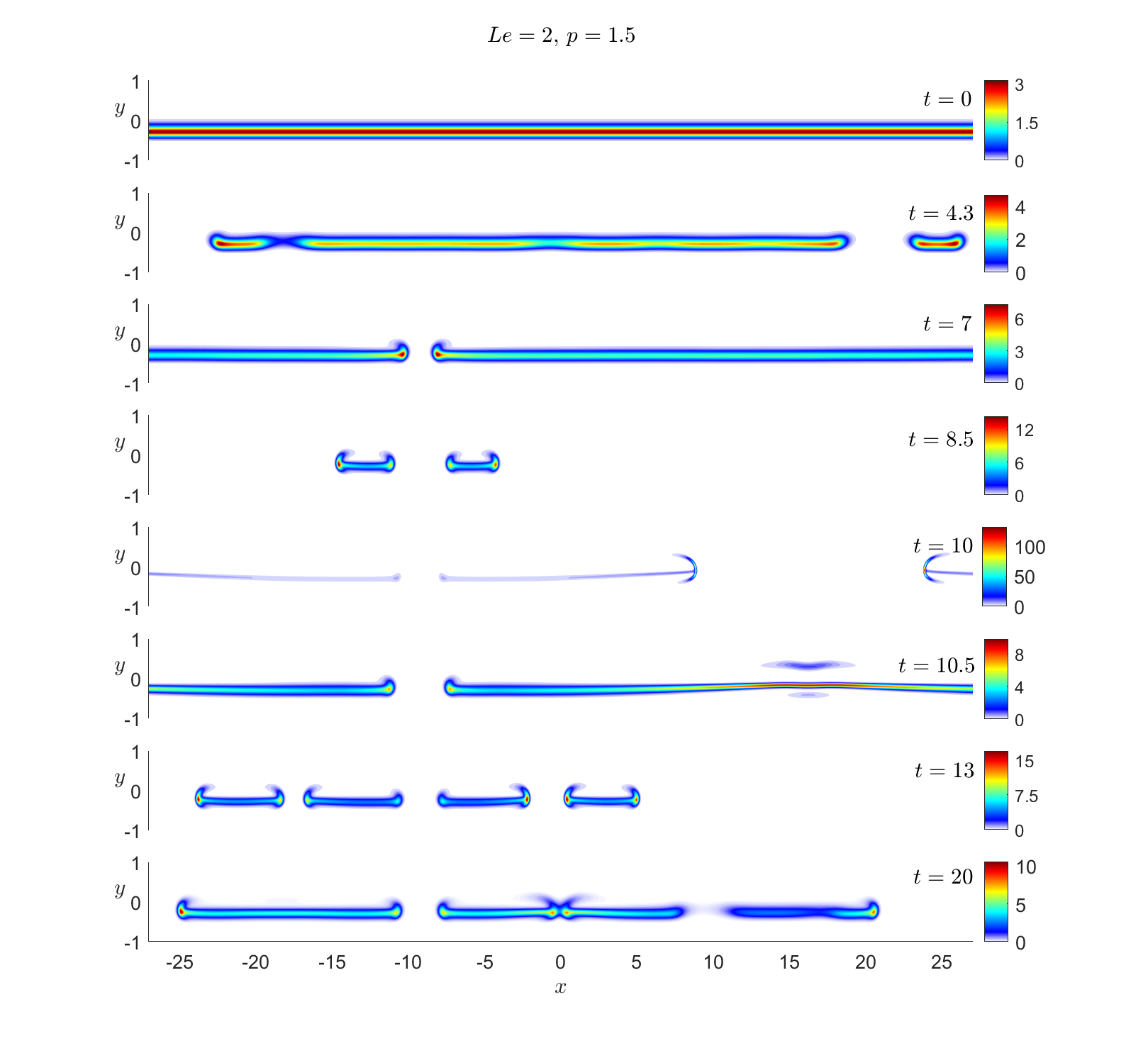}
\vspace{-1cm}
\caption{Reaction rate ($\omega$) fields at selected values of time $t$ for $\Lew=2$, $p=1.5$, $\delta=0.01$, $\beta=10$, $\alpha=0.85$ and $S=1$.} \label{fig:Le2p1.5time}
\end{figure}

In this section, we present computations for parameter values which fall in the region labelled `finite wavelength instability' in Fig.~\ref{fig:regime}. The computations starting from the initial condition corresponding to a planar diffusion flame are performed in a large number of cases, but we shall in our presentation focus on four illustrative cases. Two of these cases correspond to the points $(\Lew=1.6,p=2)$ and $(\Lew=1.6,p=3)$ in Fig.~\ref{fig:regime}; the first point is selected to be close to the boundary $p=p_A$ in Fig.~\ref{fig:regime} and the second further from this boundary. The two other cases correspond to the points $(\Lew=2,p=1.5)$ and $(\Lew=2,p=2)$ in Fig.~\ref{fig:regime}; the first point is selected to be close to the boundary $p=p_B$ in Fig.~\ref{fig:regime} and the second further from this boundary. The time evolution of the flame in these four cases are described in figures~\ref{fig:Le2p2time}-\ref{fig:Le2p1.5time}, showing  the fields of the reaction rate $\omega$ for selected values of time. Note that in these figures, the coordinate $x$ (which measures the horizontal distance scaled by the mixing layer thickness $L$) is used instead of $X=x/\sqrt{1+p^2}$ in order to better appreciate the size of the emerging cells.   %The numerical result for these four cases are shown in Fig.~\ref{fig:cells}, where the maximum temperature $\theta_{\mathrm{max}}$ is plotted versus time $t$.

We begin by commenting on the development of the instability in the simplest case which corresponds to $(\Lew=2,p=2)$, shown in Fig.~\ref{fig:Le2p2time}. Here we observe that the planar diffusion flame initially splits into a series of strongly burning regions, separated by locally quenched gaps. The strongly burning regions, each of which terminates at two \textit{edge-flames}, are found to readily adjust the gaps among themselves and settle into an apparently stable steady state. At the onset of instability occurring at $t\approx 3.9$, we note the emergence of $8$ cells or strongly burning regions. This number of cells, which is observed to persist in time, is in good agreement with that predicted
by the linear stability analysis of $\S$\ref{disp}. Indeed the maximum growth rate according to the linear analysis correspond to $\kappa=\kappa_m=1.65$, which for our horizontal domain $x=30\sqrt{1+p^2}\approx 67$ yields $7.86\approx 8$ cells. Similar observations can be made for unstable flames when $p\gg p_A$ and $p\gg p_B$ as our extensive set of computations (not shown herein) confirm. In particular, it is observed that unstable diffusion flames evolve into apparently stable steady cells and the number of these cells is rather correctly predicted by the linear stability analysis.

We now comment on the other three cases under consideration where a more complex dynamics is obtained. Figure ~\ref{fig:Le1.6p3time} illustrates the flame evolution when $(\Lew=1.6,p=3)$.
 Here after the initial splitting of the planar flame, the strongly burning regions continue to shrink  to form  \textit{flame spots} at $t\approx 15.8$. Some of these spots eventually quench, leaving a wide gap of fresh unburnt reactant mixture ($t\approx 16.7$). This wide gap is invaded by propagating edge-flames from opposite ends and gradually shrinks to disappear at $t\approx 21.5$. The resulting nearly planar flame, replacing the wide gap, becomes itself unstable and splits into a series of cells, which are found to settle into an apparently stable stationary cellular structure.

Turning now to Fig.~\ref{fig:Le1.6p2time} pertaining to the case $(\Lew=1.6,p=2)$, we observe that the development of the flame instability is roughly similar in its initial and final stages to that in the previous figure. In particular,
 an apparently stable steady structure is ultimately obtained, as in Fig.~\ref{fig:Le1.6p3time}, despite peculiar dynamics in intermediate stages.

\begin{figure}[h!]
\centering
\includegraphics[scale=0.7]{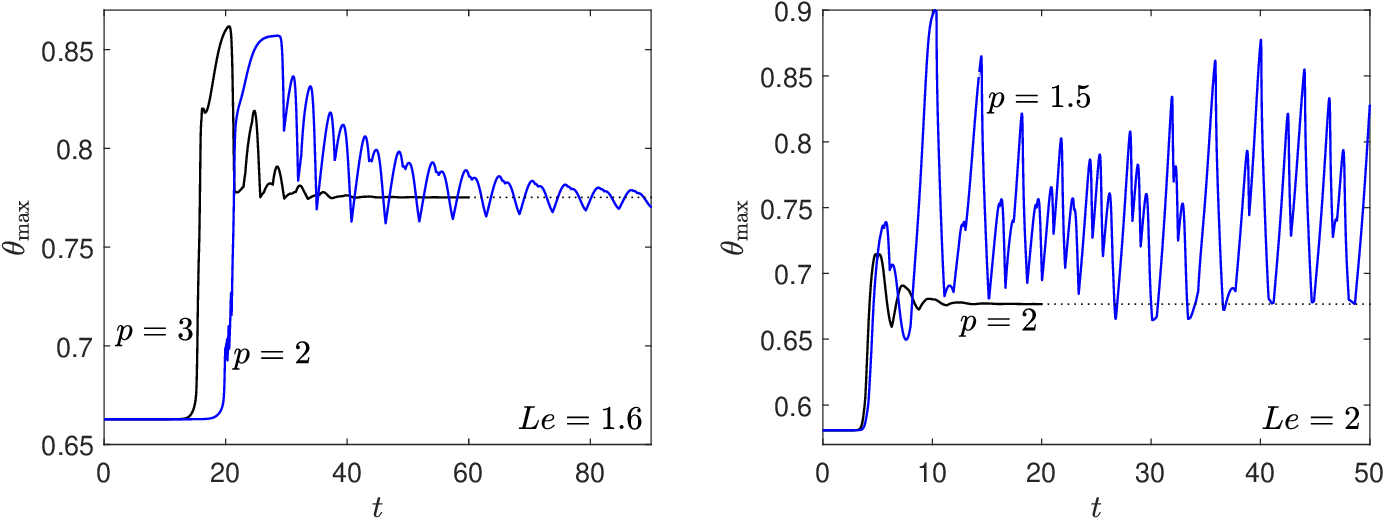}
\caption{Maximum temperature ${\theta}_{\mathrm{max}}$ as a function of time $t$, calculated for $\Lew=1.6$ (left) and $\Lew=2$ (right) with $\delta=0.01$ and selected values of $p$. All computations are carried out with $\beta=10$, $\alpha=0.85$ and $S=1$. The cases correspond to those of figures~\ref{fig:Le2p2time}-\ref{fig:Le2p1.5time}. All curves asymptote to a steady state except when $\Lew=2$, $p=1.5$.} \label{fig:cells}
\end{figure}

We now discuss the case $(\Lew=2,p=1.5)$ which is represented in Fig.~\ref{fig:Le2p1.5time}. Here the most notable new feature is that the unstable diffusion flame does not evolve into a steady state. It evolves instead into an irregular, apparently chaotic state, where several physical mechanisms are at play including edge flame propagation and merging, flame splitting, etc. The full dynamics of the flame is best appreciated by examining the animated time evolution which is included as a supplementary material. Note that the chaotic behaviour observed seems to occur when $p$ is close to $p_B$, corresponding to the continuous transitions discussed in $\S$\ref{jc}.

 Finally, to complement our discussion of figures~\ref{fig:Le2p2time}-\ref{fig:Le2p1.5time}, we plot the function ${\theta}_{\mathrm{max}}(t)$ for the cases of these figures in Fig.~\ref{fig:cells}. The figure confirms the asymptotic evolution of the unstable diffusion flame into a steady state, except for the case $(\Lew=2,p=1.5)$ of Fig.\ref{fig:Le2p1.5time}, where an irregular time dependent behaviour is obtained.

 %In all four cases, the initial increment in the maximum temperature corresponds to the shrinking of  strongly burning reaction zones. In the left subfigure corresponding to $\Lew=1.6$, the increase in the maximum temperature beyond the small kink found near $t\approx 22$ for $p=2$ and $t\approx 16$ for $p=3$ corresponds to the (almost) freely propagating triple flames (see Fig.~\ref{fig:Le1.6p3time} and~\ref{fig:Le1.6p2time}). The final steady state temperature is reached in an oscillating manner accompanied by corresponding oscillating changes in the gap widths between the cells. For the case $\Lew=2$ with $p=2$, there was no freely propagating edge flame dynamics (see Fig.~\ref{fig:Le2p2time}) so that the maximum temperature do not exhibit the corresponding increase in ${\theta}_{\mathrm{max}}$ as in the left subfigure. At last, we may note that the curve corresponding to $\Lew=2$ with $p=1.5$ reflects the chaotic nature of the nonlinear evolution near continuous transitions.

\subsection{Additional comments on unstable flames when $p\lessapprox p_A$}

Consideration of the left subfigure of Fig.~\ref{fig:delta0.01} shows that the planar diffusion flame is expected to disintegrate near instability onset into a cellular structure if $p>p_A$, as confirmed in our two-dimensional computations just presented. In fact, such cellular structures once formed can persist if $p$ is decreased to a certain extent below $p_A$, which may have some implications on the formation of diffusion flame streets in experiments. This statement can be confirmed by numerical computations (not shown) adopting as initial conditions cellular structures or an array of hot spots, as done in~\cite{carpio2020near,zhou2019effect}. For example, we have performed computations for the case $\Lew=1.6$ and $p=1.5$, for which the most unstable mode corresponds to the oscillatory longwave mode ($\kappa=0$), according to  Fig.~\ref{fig:delta0.01} (left), although the finite wavelength mode is also unstable in this case. It is found that if the initial condition is taken to be a cellular structure (say, the steady solution corresponding to $\Lew=1.6$ and $p=3$), then this structure evolves into another stable cellular structure. In contrast, if the initial condition is taken to be the planar diffusion flame solution, then an oscillatory longwave instability leading to flame quenching is observed.

\section{Concluding remarks}

In this paper, we have examined the stability of a planar diffusion flame aligned with the direction of a shear flow within a simple narrow channel model. Particular attention has been focused on the effect of the flow Peclet number on flame stability. A linear stability analysis, supported by time dependent numerical simulations, confirms that the diffusion flame can be unstable for $\Lew$ sufficiently larger than one and that the nature of the emerging instability depends critically on the (scaled) Peclet number $p$. Specifically, it is shown that a longwave instability with or without time oscillations is obtained for small values of $p$, while a finite wavelength (cellular) instability is obtained for $p$ above a critical value $p_c$. The instability domains are clearly delimited for the illustrative examples of Fig.~\ref{fig:regime} and~\ref{fig:regime2}, where $p_c$ corresponds to the curves labelled $p_A$ (jump transition) and $p_B$ (continuous transition). The longwave instability is found to typically lead to flame quenching. The more crucial finding is however that a critical value $p_c$ of the Peclet number is shown to exist, above which the instability is cellular;  this is consistent with the experimental observations made in the original studies~\cite{miesse2005experimental,miesse2005diffusion}. Our findings provide an original explanation to the issues related to the two questions raised in the introduction, namely, the existence of a critical Peclet number and the disintegration of the planar flame when $\Lew>1$. Our explanation is largely based on Taylor's dispersion mechanism which leads to the presence of effectively two distinct Lewis numbers, $\Lew_x$ and $\Lew$, in the longitudinal and transverse directions respectively, as discussed in the introduction. The necessary condition for the occurrence of the finite wavelength (or cellular) instability is found to be $\Lew_x<1$ provided the fuel Lewis number $\Lew>1$.

%In summary, the two questions raised in the introduction in connection with the experimental observations made in non-premixed microcombustion devices are answered through our linear stability analysis and time dependent numerical simulations of a planar diffusion flame aligned with the shear flow. The underlying mechanism that results cellular structures in heavy fuels thus may be attributed to the reversal of roles between a strongly diffusing reactant and a weakly diffusing reactant, generated by Taylor's dispersion mechanism.
To close this section, we make a few remarks regarding possible extensions of the current study. The first useful extension is to account for heat losses which were deliberately neglected herein in order to isolate the effect of Taylor dispersion as being on its own a driving mechanism of the cellular flame instability. Heat loss effects can be significant, however, in experiments such as those involving methane diffusion flames~\cite{miesse2005experimental,miesse2005diffusion}. Such diffusion flames, characterised by a (fuel) Lewis number close to unity, are expected to be stable according to our adiabatic investigation, but are found to disintegrate into flame streets in practice, which may be explained by accounting for heat losses~\cite{mohan2017diffusion}. Another useful extension of this study is to account for the influences of stoichiometry by considering values of $S\neq 1$ since flammability limits can be greatly extended due to the formation of stable cellular structures in the range $\delta<0$ when $S\ll 1$~\cite{zhou2019effect,carpio2020near}. It would also be interesting to have experimental verification of some key findings of the current investigation such as the existence of cellular structures near the jump transition at $p_A$ identified above, both for $p\lessapprox p_A$ and $p>p_A$. \blue{Finally, it is worth investigating whether the cellular instability for large Lewis numbers induced by Taylor dispersion identified herein for diffusion flames can also be found for premixed flames aligned with the direction of a shear flow. Such investigations would enrich our appreciation of premixed flame instabilities in confined systems considered in recent studies~\cite{veiga2020unexpected,martinez2019role,gu2021propagation,dominguez2022stable}.}

\section*{Acknowledgements}
This work was supported by the UK EPSRC through grant EP/V004840/1.

\section*{Appendix A: Stability of the Burke-Schumann flame}
In this appendix, we solve the linear eigenvalue problem~\eqref{H eig}-\eqref{BC eig} in the Burke-Schumann limit $\delta\rightarrow \infty$. In this limit, reactants leakage across the reaction sheet is zero, so that the perturbations are only associated with the diffusion process in the outer zones, which are in chemical equilibrium, characterised by $f=g=0$. Then, equation~\eqref{YF eig} can be solved to yield
\begin{equation}
    \sigma_n= -(4\pi^2 n^2/\Lew + \kappa^2/\Lew_x), \quad \widetilde{Y}_{F,n} = \sin(2\pi n y) \quad \text{where} \quad n=1,2,3,\dots
\end{equation}
indicating that $\sigma_n<0$ and hence the flame is unconditionally stable for all wavelengths (see also~\cite[p. 238]{kim1996diffusional}). The function $\widetilde{H}_{n}$ can be obtained explicitly by solving
\begin{equation}
    \left(\frac{d^2}{dy^2} - \kappa^2 + \frac{4\pi^2n^2}{\Lew} + \frac{\kappa^2}{\Lew_x}\right)\widetilde{H}_n = - \left[\frac{(\Lew-1)}{\Lew}4\pi^2n^2 - \frac{(\Lew_x-1)}{\Lew_x} \kappa^2\right] \sin(2\pi ny)
\end{equation}
subject to $\widetilde{H}_{n}(-1)=\widetilde{H}_{n}(+1)=0$.

\bibliographystyle{elsarticle-num}

\bibliography{sample}

\end{document}